\documentclass[conference]{IEEEtran}
\IEEEoverridecommandlockouts
\usepackage{amssymb,amsfonts}
\usepackage{algorithmic}
\usepackage{graphicx}
\usepackage{stfloats}
\usepackage{textcomp}
\usepackage{xcolor}
\usepackage{epsfig}
\usepackage{includes/citesort}
\usepackage{amssymb}
\usepackage{amsmath}
\usepackage{color}
\usepackage{url}
\usepackage{subcaption}
\usepackage{cite}
\usepackage{mathtools}
\usepackage{xspace}
\usepackage{comment}
\usepackage{array}
\usepackage{hyperref}
\usepackage{multicol}
\usepackage{lipsum}
\usepackage{inputenc}
\usepackage{booktabs}

\newcolumntype{P}[1]{>{\centering\arraybackslash}p{#1}}

\newlength{\figurewidth}
\newlength{\smallfigurewidth}
\makeatletter
\def\smallunderbrace#1{\mathop{\vtop{\m@th\ialign{##\crcr
   $\hfil\displaystyle{#1}\hfil$\crcr
   \noalign{\kern3\p@\nointerlineskip}%
   \tiny\upbracefill\crcr\noalign{\kern3\p@}}}}\limits}
\makeatother

\def\BibTeX{{\rm B\kern-.05em{\sc i\kern-.025em b}\kern-.08em
    T\kern-.1667em\lower.7ex\hbox{E}\kern-.125emX}}
    
\begin{document}

\title{Value-Compressed Sparse Column (VCSC): Sparse Matrix Storage for Single-cell Omics Data}

\author{
\IEEEauthorblockN{Seth Wolfgang}
\IEEEauthorblockA{\textit{Department of Intelligent Systems Engineering}\\
\textit{University of Indiana}\\
Bloomington, IN, USA\\
seawolfg@iu.edu}

\and 

\IEEEauthorblockN{Skyler Ruiter}
\IEEEauthorblockA{\textit{Department of Intelligent Systems Engineering}\\\textit{University of Indiana}\\
Bloomington, IN, USA\\
sruiter@iu.edu}

\and

\IEEEauthorblockN{Marc Tunnell}
\IEEEauthorblockA{\textit{Department of Computer Science}\\
\textit{Purdue University}\\
West Lafayette, IN, USA\\
mtunnell@purdue.edu}

\and

\IEEEauthorblockN{Timothy Triche Jr.}
\IEEEauthorblockA{\textit{Department of Epigenetics}\\
\textit{Van Andel Institute}\\
Grand Rapids, MI, USA\\
tim.triche@vai.org}

\and

\IEEEauthorblockN{Erin Carrier}
\IEEEauthorblockA{\textit{College of Computing}\\
\textit{Grand Valley State University}\\
Allendale, MI, USA\\
carrieer@gvsu.edu}

\and

\IEEEauthorblockN{Zachary DeBruine}
\IEEEauthorblockA{\textit{College of Computing}\\
\textit{Grand Valley State University}\\
Allendale, MI, USA\\
debruinz@gvsu.edu}

}

\maketitle

\begin{abstract}
Genomics datasets, such as single-cell transcriptomics, are often very large and highly sparse, posing significant challenges for both storage and computation. As the scale of data generation accelerates, efficiently compressing these datasets becomes crucial. Current compression methods, like the popular Compressed Sparse Column (CSC) format, capitalize only on sparsity but overlook other properties like redundancy, which can offer additional opportunities for compression. Genomics data, especially single-cell assays, often exhibit high redundancy within columns, making traditional sparse formats inefficient for in-core computation. In this paper, we present two extensions to CSC: (1) Value-Compressed Sparse Column (VCSC) and (2) Index- and Value-Compressed Sparse Column (IVCSC).
VCSC takes advantage of high redundancy within a column to further compress data up 1.9-fold over CSC on real data, without significant negative impact to performance characteristics.
IVCSC extends VCSC by compressing index arrays through delta encoding and byte-packing, achieving up to a 4.4-fold decrease in memory usage over CSC on real data.
Our benchmarks show that VCSC and IVCSC can be used in compressed form with little added computational cost. These formats represent a step forward in balancing the growing demands of data storage and processing in the era of large-scale genomics.\end{abstract}

\begin{IEEEkeywords}
sparse data, matrix compression, data redundancy
\end{IEEEkeywords}

\section{Introduction}
Sparse data is mostly zero or missing, and is often encoded in sparse matrices that avoid explicit storage of these values.
Sparse matrices are abundant in many domains that involve scientific computing, machine learning, and data engineering. 
In these domains, software priorities are often a combination of memory usage and fast compute, with these goals usually being at odds with one another.

\begin{figure*}
  \centering
  \includegraphics[width=0.7\textwidth]{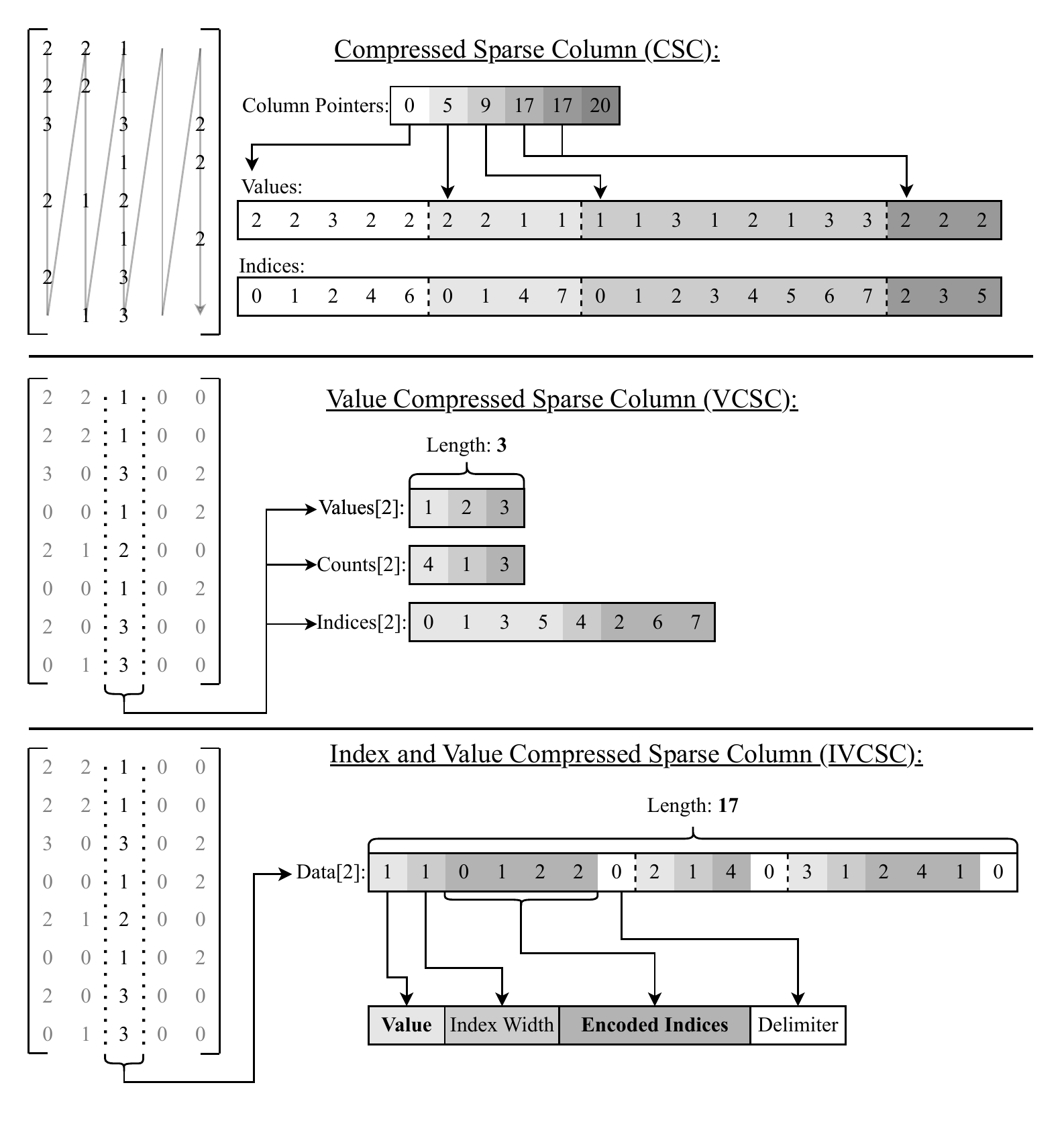}
  \caption{Visual comparison of sparse matrix storage formats.}\label{fig:format_diagram}
\end{figure*}

General purpose sparse matrix formats, such as Coordinate (COO) or Compressed Sparse Column (CSC), offer compression with reasonably fast compute. 
Specifically, COO stores the matrix in triplet format, storing a row-index, column-index, and value for each nonzero value.
As depicted in \autoref{fig:format_diagram}, for each nonzero value, CSC (CSR) stores the value and the row-index (column-index), along with the pointer offsets to the start of each column (row).
While popular, COO, CSC, and CSR fail to leverage specific characteristics of the data, such as significant redundancy in nonzero values, common in count data and discrete distributions. 

Massive sparse datasets such as those used in genomics, often contain highly redundant values.
Take, for example, an $18082\times897734$, single-cell transcriptomics dataset from 10x Genomics containing the counts of genes in single-cells \cite{genomics_lung_cancer}.
This matrix is 92\% sparse and contains approximately 1.3 billion nonzero values.
Despite the large number of nonzero values, there are only about $7,000$ unique values.
Despite being highly redundant, as shown in \autoref{tab:tableData}, in CSC format this dataset requires roughly 7GB ($\approx 24\%$ of the dense storage required).
As genomics data continues to grow, memory limitations will become increasingly prohibitive to in-core analysis and massive data integration.

In this paper, we introduce two novel sparse matrix formats that aim to address this limitation of conventional sparse matrix formats on highly redundant data: (1) Value-Compressed Sparse Column (VCSC) and (2) Index- and Value-Compressed Sparse Column (IVCSC).
Using VCSC, the size of the previously mentioned 10x Genomics dataset is reduced to $68\%$ of CSC storage.
IVCSC enables further compression, reducing the size to $23\%$ of CSC storage.
These formats are implemented in C++ and available on Github (\url{https://github.com/Seth-Wolfgang/IVSparse}) and archived on Zenodo (\url{https://zenodo.org/doi/10.5281/zenodo.10084031})\cite{wolfgang2023}.

This paper is organized as follows:
Related work is discussed in section 2, methods underlying VCSC and IVCSC are discussed in section 3, experimental results are presented in section 4, and a conclusion and discussion are given in section 5.
    

\section{Related Work}
A variety of compressed matrix formats exist, for both sparse and dense matrices, with most formats focusing on enabling fast compute.
As is typical with sparse matrix formats, we limit our discussion of related work to only lossless compressed matrix formats.

\subsection{Sparse Matrix Compression}
A popular approach is to create a data structure that groups data into blocks \cite{Borštnik2014, buluç2009, pinar99, karakasis2013}.  Block storage formats often take advantage of natural patterns in the data to speed up Sparse Matrix-Vector Multiplication (SpMV), utilizing this structure for better cache access \cite{buluç2009, pinar99}, optimizing for distributed systems \cite{Borštnik2014}, or for better memory bandwidth via data compression \cite{karakasis2013}.

The authors of \cite{willcock2006} utilize a form of data compression aimed at achieving faster SpMV computations.  Specifically, index data is delta-encoded and decompressed to enable faster computation on sparse data with distinct local density patterns \cite{willcock2006}.

In \cite{kourtis2008}, two modifications to further compress CSR, namely CSR Delta Unit (CSR-DU) and CSR Value Indexed (CSR-VI) are presented.
CSR-DU compresses the column indices and row pointers in CSR by breaking up the matrix into units, applying delta encoding over each unit with flags to store when a unit starts a new row.
CSR-VI compresses values by storing unique values in a lookup table, reducing the memory footprint for redundant data but necessitating the storage of references for each nonzero value.
While similar in idea to our approach, \cite{kourtis2008} attempts to optimize SpMV performance through compression, whereas we focus on optimizing compression with limited performance impact on common operations.

\subsection{Dense Matrix Compression}
While most existing sparse matrix compression formats fail to take advantage of redundancy, many dense data compression algorithms are designed to capitalize on redundancy.
For example, Huffman encoding constructs a dictionary of keys where more common symbols have smaller keys \cite{huffman52}.
Lempel-Ziv-Welch (LZW) \cite{lzw84} compression is similar to Huffman, but targets redundancy by compressing repetitive patterns.
LZW excels on data where repeated patterns are prevalent, albeit at the cost of increased storage for unique patterns.
Run-length Encoding (RLE) encodes consecutive identical symbols as a single count-value pair, and the compression ratio is directly correlated with the extent of consecutive redundant values in the data.

\section{Methods} 

\subsection{Value-Compressed Sparse Column (VCSC) Format}
VCSC takes advantage of the per-column redundancy of values by storing each unique value only once per column in which it occurs.
For each column, VCSC stores three arrays, one for unique values, one for value counts, and one for row indices.  
The entries of the values array are the unique values in the column.
The corresponding entries of the counts array are the number of times each value occurs in the column.  
The entries of the indices array are the row indices of each occurrence.  
These row indices are ordered first by value and within each value ordered in increasing order of row index.  
An example of VCSC format for a single column of a sparse matrix is shown in the middle panel of \autoref{fig:format_diagram}.

By reordering the row indices first by value, we eliminate the need to store references to each element, as is necessary in CSC-VI.  
While ordering first by value, then by row index significantly improves compression for highly redundant data, traversal through the rows in a column is no longer guaranteed to be performed sequentially.

To evaluate the memory usage of VCSC, we first consider the memory usage of CSC which is given by
\begin{align}
\label{eq:cscmem}
	\text{CSC}_\text{size} = & \underbrace{val_\text{size}*nnz}_{\text{bytes for nonzero vals}} + \underbrace{idx_\text{size} * nnz}_\text{bytes for indices} \notag \\
	& \hspace{3cm} + \underbrace{idx_\text{size}*(ncol+1)}_\text{bytes for col pointers},
\end{align}
where $nnz$ is the number of nonzeros and $val_\text{size}$ and $idx_\text{size}$ are the byte sizes for values and indices, respectively.
In contrast, the memory usage of VCSC is given by
\begin{align}
\label{eq:vcscmem}
	\text{VCSC}_\text{size} = & \sum_{i=1}^{nCols} \left( \underbrace{val_\text{size} * nUniq_i}_\text{bytes for values} + \underbrace{idx_\text{size} * nUniq_i}_\text{bytes for counts} \right. \notag \\
	& \hspace{2cm} \left. + \underbrace{idx_\text{size} * nnz_i}_\text{bytes for indices} + \underbrace{idx_\text{size}}_\text{len} \right),
\end{align}
where $nUniq_i$ is the number of unique values in column $i$, and $nnz_i$ is the number of nonzeros in column $i$.
Unlike CSC (and CSC-VI), the only component of \autoref{eq:vcscmem} which grows at a fixed rate with the number of nonzeros in a column is the memory for the indices.

\subsection{Index- and Value-Compressed Sparse Column (IVCSC) Format}

Whereas VCSC compresses just values, IVCSC further compresses VCSC by also compressing the indices.
For each column, IVCSC stores a single array.  
This array contains sections, where each section is a unique value, followed by the index width, followed by the row indices where that value occurs, followed by a delimiter (a zero) to indicate the end of the row indices for that value.
Within a single unique value, the indices are compressed by positive-delta encoding, as shown in the bottom pane of \autoref{fig:format_diagram}, and then byte-packed.

By positive-delta encoding the row indices, the magnitude of the stored indices is reduced.  
Byte-packing the encoded indices discards leading bytes that are zero, further reducing the storage required for each index, while still allowing efficient traversal through the indices (which would not be the case with bit-packing).
Depending on the redundancy and density of the data, it is often possible to represent indices for frequently occurring values with a single byte.

As with VCSC, traversal through the rows in a column is not necessarily sequential.
Furthermore, traversal through a column in IVCSC requires decoding the positive-delta-encoded indices and checking for end-of-value delimiters.

The memory usage of IVCSC is given by
\begin{align}
\label{eq:ivcscmem}
\text{IVCSC}_\text{size} = & \sum_{i=1}^{nCols} \left( \underbrace{8}_\text{len} + \underbrace{nUniq_i * \left(val_\text{size} + 1\right)}_\text{bytes for value and idx width} \right. \notag \\
& \hspace{1cm} \left. + \underbrace{\sum_{j=1}^{nUniq_i}{\left(nnz_j + 1\right) * idxWid_j}}_\text{bytes for encoded indices and delim} \right),
\end{align}
where the 8 bytes (denoted by $\texttt{len}$) are used to store the size of the data array, $nnz_j$ is the number of times the unique value $j$ appears in the column, and $idxWid_j$ is the byte width of the unique value's indices after positive-delta encoding and byte-packing. 
Comparing \autoref{eq:vcscmem} and \autoref{eq:ivcscmem}, one can see the bytes for storing the value data is similar, with the main difference being that the counts of unique values are replaced with delimiters, which are often slightly smaller.
Index compression is apparent in that $idxWid_j$ is potentially different for each unique value in the column.

IVCSC is suitable for very large, highly redundant matrices.
Compared to VCSC, IVCSC further prioritizes compression at increased computation and traversal time.

\begin{figure*}
    \centering
    \includegraphics[width=\textwidth]{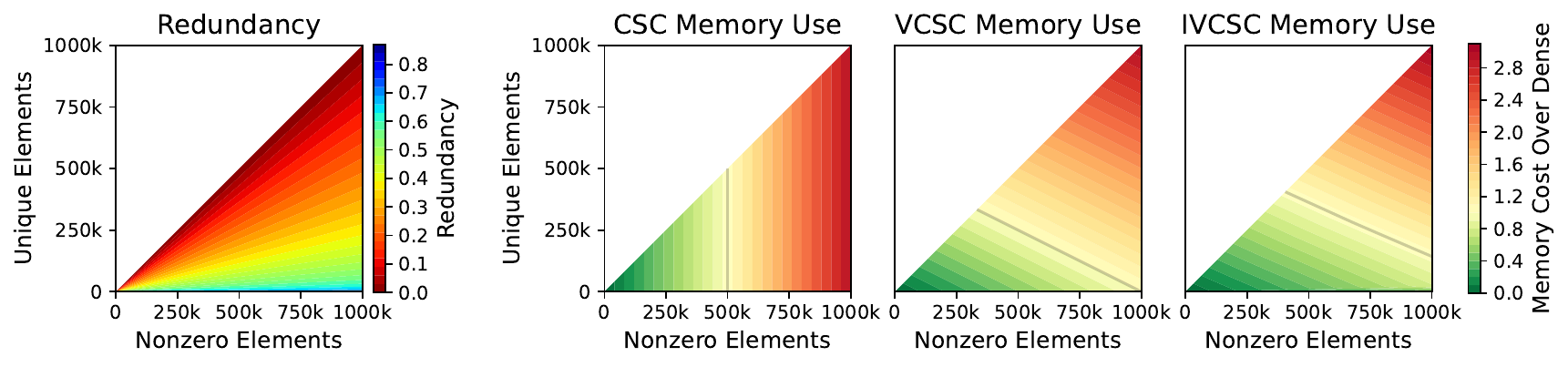}
    \caption{Visualization of redundancy and memory cost for varying numbers of unique elements and nonzeros.  The leftmost graph shows redundancy as a function of both unique elements and nonzero elements. The right three graphs show CSC, VCSC, and IVCSC memory usage (as a ratio over dense) for varying numbers of unique elements and nonzeros}
    \label{fig:redundancy-memory-plot}
\end{figure*}

\section{Experimental Results}

\subsection{Setup and Implementation}

All benchmarking is performed on large memory (1.5TB RAM) nodes on a dedicated cluster, with Intel(R) Xeon(R) Gold 6226 processors.
Each node has 4 sockets, with each CPU having a max clock speed of 2.70GHz, 12 cores, 1.5MB L1d cache (per core), 1.5MB L1i cache (per core), 48MB L2 cache (per core), and 77MB L3 cache (per CPU).
Each node has 8 NUMA region, and, to limit any potential impact of the order in which the formats are run, the benchmarking runs for each storage format are bound to a specific NUMA region.
At the time of benchmarking, this cluster runs RHEL 9.2 and all programs were compiled using GCC version 11.2.0 with the flags \texttt{-O3} and \texttt{--std=c++17}.
OpenMP was disabled for all benchmarking.

We evaluate the performance with respect to both compression ratio and computational performance.  Compression ratio is evaluated on both randomly generated matrices and a variety of datasets that represent a range of real-world use cases.

Computational performance is evaluated on a variety of common operations.
For computational performance, we benchmark against the CSC implementation of the Eigen C++ library, a widely used matrix and linear algebra package \cite{eigen}.
Each timing benchmark is given two cold starts, which are not timed.  All timing is performed using  \texttt{clock()} in the C \texttt{time.h} header file, with results reported as the mean of five timed repeats.  To ensure no operations are optimized out, results for each repeat are accumulated in  \texttt{volatile} memory.
To isolate as many variables as possible and because we benchmark the fundamental data structure, not our implementation of BLAS-like routines, we implement the same naive algorithm for SpMV and SpMM on the Eigen CSC matrix and our implementations.   

For all matrices, indices are stored as 4-byte integers (excluding any positive-delta encoding and byte-packing).  For simulated matrices, values are stored as single precision (4-byte) floats. For real datasets, values are stored as the smallest typically available type appropriate for the values.

\subsection{Memory Usage}

In order to quantify the efficiency of our format on redundant data, for all columns with nonzero elements we define the redundancy of the $i$-th column as
\begin{equation}
\label{eq:redundancy}
    r_i = 1 - \frac{1}{\log_{10}\left(nnz_i\right) - \log_{10}\left(nUniq_i\right) + 1}.
\end{equation}

\newcommand{\name}{MMR\xspace}

This metric captures the magnitude of the difference between the number of non-zero elements and the number of unique elements.
While \autoref{eq:redundancy} differs from the natural first instinct of how to measure redundancy (which most would consider to be $1-nUniq_i/nnz_i$), it has far more resolution for describing highly redundant data.
For instance, a very large dataset with 1M  nonzeros per column and 100k unique values per column would not naturally be considered highly redundant, but would be considered highly redundant by the natural instinct for redundancy.
In contrast, our metric for redundancy captures the relative scales of the number of nonzeros compared to the number of unique values.
Redundancy, $r_i$, is averaged over all columns with nonzero elements, giving the mean matrix redundancy (\name{}).

\autoref{fig:redundancy-memory-plot} shows redundancy as a function of the number of nonzero elements in a given column and unique elements in the form of a heatmap.
This demonstrates that high redundancy requires a small number of unique elements, even for a matrix with a large number of rows.
Additionally, \autoref{fig:redundancy-memory-plot} compares the memory use for CSC, VCSC, and IVCSC, with grey lines denoting a sparse-to-dense memory ratio of 1 (e.g. no memory savings).  
While all of CSC, VCSC, and IVCSC are able to achieve significant compression for matrices with few nonzero elements, both VCSC and IVCSC are able to compress further when there are relatively few unique values.
Furthermore, at very high levels of redundancy, both VCSC and IVCSC demonstrate the ability to compress matrices that would otherwise be considered too dense to benefit from a sparse matrix format, including fully dense matrices.

\begin{figure*}
\centering  
\begin{subfigure}[t]{0.47\textwidth}
\centering
    \includegraphics[width=.8\textwidth]{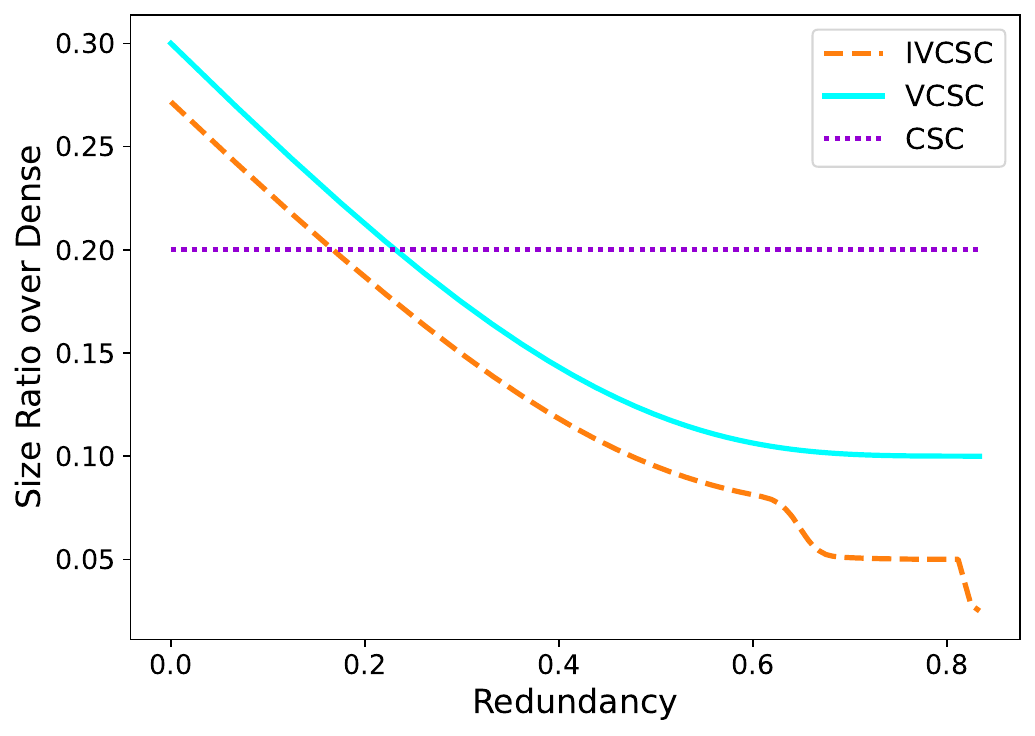}
    \caption{Comparison of size (as ratio over dense) vs redundancy for various sparse storage formats on random $1 \text{ million} \times 25$ matrix.  }  
    \label{fig:size_plot.1} 
\end{subfigure}
\hspace*{4pt}
\centering
\begin{subfigure}[t]{0.47\textwidth}
\centering
    \includegraphics[width=.8\textwidth]{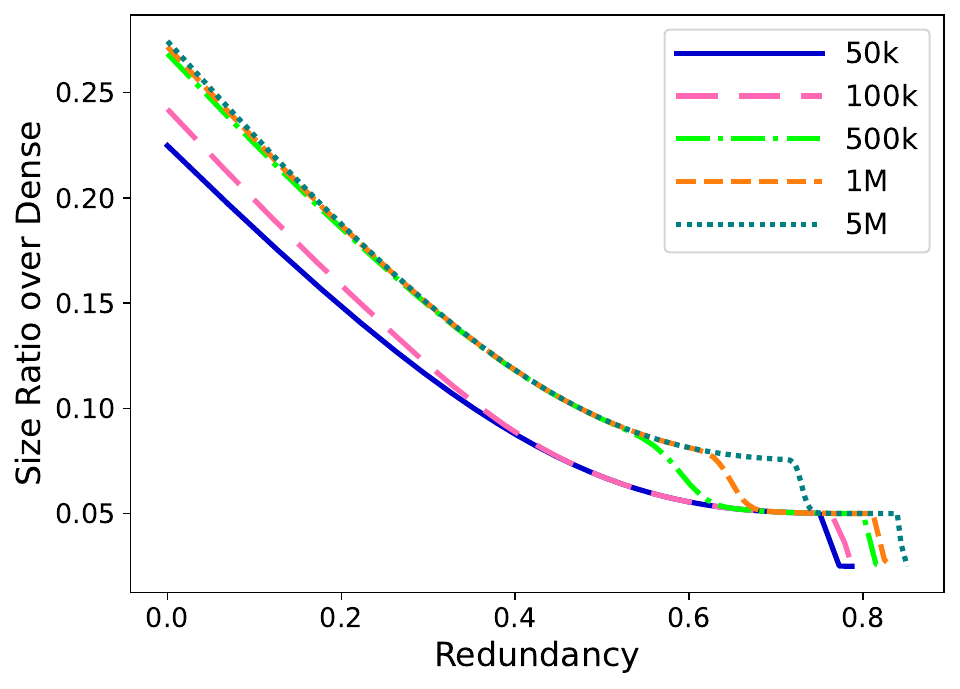}
    \caption{IVCSC size (as ratio over dense) for random matrices with varying numbers of rows and a fixed 25 columns.}
    \label{fig:size_plot.2}
\end{subfigure}
\caption{Memory usage comparison on randomly generated matrices with 90\% sparsity, with values stored as floats (4 bytes) and indices stored as 4 byte integers.}
\label{fig:size_plot}
\end{figure*}
  
\autoref{fig:size_plot.1} shows the compression ratio over dense of CSC, VCSC, and IVCSC  as a function of \name{} on a matrix with $90\%$ sparsity.
For \autoref{fig:size_plot.1}, a single matrix is generated at the beginning of a session, the values of which are then modified to change the redundancy in further runs.
Regardless of redundancy, CSC requires $20\%$ of the dense memory.
\autoref{fig:size_plot.1} shows that for low redundancy, VCSC and IVCSC require more memory than CSC, but they are able to compress more than CSC at an MMR greater than 0.23 and 0.167, respectively.
As the matrix becomes fully redundant, VCSC asymptotically approaches a theoretical minimum of storing a single unique value per column.
The dropoffs for IVCSC represent the transition in the number of bytes in which the indices are byte-packed. 
\autoref{fig:size_plot.2} compares the behavior for IVCSC with varying numbers of rows.  
This demonstrates that the size influences what byte-packing is possible and at what redundancies byte-packing occurs.  
Furthermore, the larger the number of rows, the higher the maximum possible redundancy (which is bounded by 1 in the limit).

We test the compression on five real world datasets which represent a wide range of use cases.  These datasets are of varying redundancy and of varying types of values, both of which influence the potential savings of VCSC and IVCSC.
For each of these datasets, the name, dimensions, number of nonzeros, sparsity, \name{}, value size (determined by value type), and the resulting memory usage ratios over CSC for VCSC and IVCSC are given in \autoref{tab:tableData}.

Single-cell is a single-cell transcriptomics dataset from \cite{genomics_lung_cancer}.  It is representative of data that follows a zero-inflated negative binomial counts distribution.
\autoref{tab:tableData} shows that VCSC and IVCSC provide substantial compression compared to CSC (approximately 1.5-fold and 4.4-fold, respectively).

\begin{table*}[bp]
\centering
\caption{\label{tab:tableData}%
Memory usage comparison of CSC, VCSC, and IVCSC on real datasets from various disciplines.  All size ratios are given as percentage of CSC size.}
{

\renewcommand{\baselinestretch}{1.2}\small
\resizebox{1\textwidth}{!}
{
\begin{tabular}{|P{2.02cm}|c|c|c|c|P{0.7cm}|P{0.95cm}|P{0.7cm}|P{0.8cm}|P{1.0cm}|P{0.95cm}|P{1.0cm}|} 
\hline
\textbf{Dataset} &
\textbf{Dimensions} & 
\textbf{NNZ} & 
\textbf{Sparsity} &
\textbf{\name{}} &
\textbf{Value Size (B)} &
\textbf{Dense Size (GB)} &
\textbf{CSC Size (GB)} &
\textbf{VCSC Size (GB)} &
\textbf{VCSC Ratio} &
\textbf{IVCSC Size (GB)} &
\textbf{IVCSC Ratio} \\
\hline
Single-cell (NSCLC) & $18082\times897733$ & $1.30$e$9$ & 91.973\% & 0.66 & 2  & 30.24  & 7.28 & 4.93 & 67.78\% & 1.64 & 22.56\%  \\
\hline
Web of Science & $46985\times124836$ & 5.41e6 & 99.908\% & 0.15 & 1  & 5.46  &  0.026 & 0.022 & 84.54\% & 0.0096 & 37.10\%  \\
\hline
MovieLens & $162541\times59047$ & 2.50e7 & 97.740\% & 0.23 & 4 &  35.75  & 0.19 & 0.097 &  51.21\% & 0.046 &  24.21\% \\ 
\hline
PR02R & $161070\times161070$ & 8.19e6 & 99.968\% & 0.0024 & 8 & 193.29 &  0.092 & 0.12 & 132.50\% & 0.10 & 118.17\% \\
\hline
com-Orkut & $3072441 \times 3072441$ & 2.34e8 & 99.998\% &  0.59 & 1 & 8791.59 & 1.10 & 0.90 & 81.5\% & 0.67 & 60.80\% \\
\hline
\end{tabular}
}
}
\end{table*}

For a more detailed view of how VCSC and IVCSC compress single-cell data, we refer the reader to \autoref{tab2}, which presents the compression results for six single-cell datasets \cite{genomics_lung_cancer, NSCLC_DTC, PBMC, ColonCancer, MouseSplenocytes, JurkatCells}. In this table, VCSC consistently achieves compression to approximately 70\% of the CSC size, while IVCSC compresses down to between 23\% and 37\% of the CSC size.

Web of Science is obtained from \cite{WOSDataSet}, and processed using \texttt{CountVectorizer} with default parameters \cite{pedregosa_2011}, as a representative example of a bag-of-words model.
\autoref{tab:tableData} shows that VCSC and IVCSC both compress further than CSC (approximately 1.2-fold and 2.8-fold, respectively).  As the values only require a single byte, VCSC has limited improvement in compression ratio compared to CSC, whereas IVCSC gains significantly more from positive-delta encoding and byte-packing.

The MovieLens 25M dataset, obtained from \cite{movieLens}, is a representative example of discrete, ordinal data, containing 10 possible ratings between 0 and 5.
\autoref{tab:tableData} shows that VCSC and IVCSC provide substantial compression compared to CSC (approximately 1.9-fold and 4-fold, respectively).

The PB02r dataset, obtained from SuiteSparse \cite{Davis2011}, is representative of performance on computational fluid dynamics simulation matrices.
\autoref{tab:tableData} shows that VCSC and IVCSC perform poorly on data with low redundancy, actually increasing the storage beyond that required for CSC.

The com-Orkut dataset is a social networking dataset created in \cite{yang2012}, obtained from SuiteSparse \cite{Davis2011}.  As it represents connections between users, it is a binary matrix.  
\autoref{tab:tableData} shows that VCSC and IVCSC both compress further than CSC (approximately 1.3-fold and 1.7-fold, respectively).
As it is extremely sparse and the values require only a single byte, both VCSC and IVCSC have limited improvement over CSC.

\begin{table*}[h]
\caption{Memory usage comparison of CSC, VCSC, and IVCSC on six single-cell datasets.  All size ratios are given as percentage of CSC size.\label{tab2}}

\renewcommand{\baselinestretch}{1.2}\small
\resizebox{1\textwidth}{!}
{\begin{tabular}{|l|l|c|P{1.4cm}|P{1.4cm}|P{1.4cm}|c|P{1.4cm}|c|}
\hline%
\textbf{Dataset} & \textbf{Assay} & \textbf{Dimensions} & \textbf{Dense Size (GB)} & \textbf{CSC Size (GB)} & \textbf{VCSC Size (GB)} & \textbf{VCSC Ratio} & \textbf{IVCSC Size (GB)} & \textbf{IVCSC Ratio} \\
\hline
NSCLC & Cell Ranger 7.1.0 & $18082\times897733$ & 30.24  & 7.28 & 4.93 & 67.78\% & 1.64 & 22.56\%  \\ 
\hline
NSCLC DTCs &  Cell Ranger 6.1.0  &  $36622\times4387768$  &  321.4  &  0.978  & 0.718 &  73.38\% & 0.342 & 34.99\% \\ 
\hline
PBMCs &  Cell Ranger 8.0.0  &  $38616\times1899847$  &  146.7   &  0.332 &  0.243 & 73.17\% & 0.116 & 34.94\% \\ 
\hline
Human Colon Cancer  &  Space Ranger 2.1.0  &  $37082\times 14284$  &  1.059  &  0.134 & 0.0908 &  67.78\% & 0.0366 & 27.29\% \\ 
\hline
Mouse Splenocytes  &  Cell Ranger 7.1.0  &  $32344\times9430818$  &  610.1 &  2.514 & 1.779 &  70.76\% & 0.919 & 36.57\% \\ 
\hline
Jurkat Cells  &  Cell Ranger 7.0.0  &  $37143\times 9489059$  &  704.9  & 5.192 & 3.57 & 68.68\% & 1.49 & 28.74\% \\ 


\hline
\end{tabular}
}
\end{table*}

\subsection{Computational Performance}

\begin{figure*}
\centering  
\hspace*{1cm}
\begin{subfigure}[t]{0.33\textwidth}
\includegraphics[width=\textwidth]{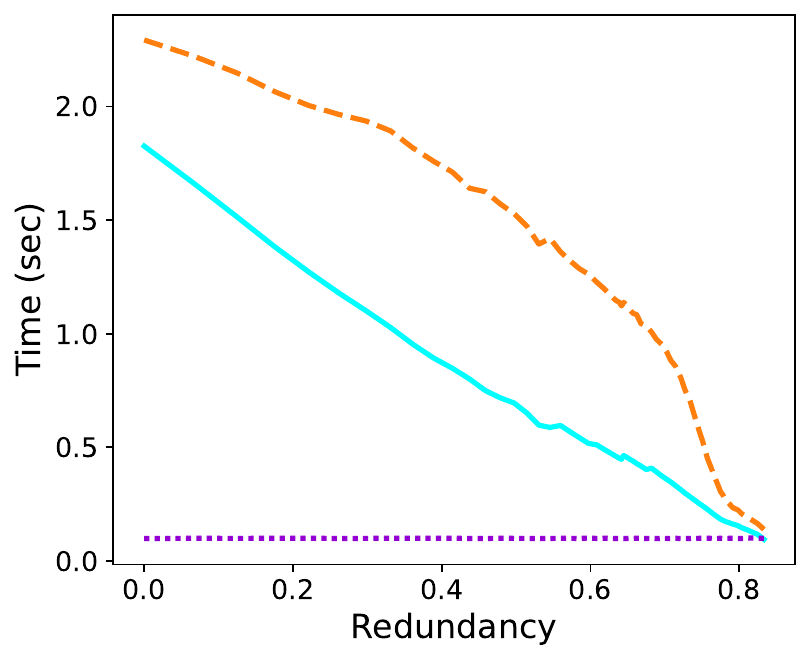}
\caption{Constructor}\label{fig:construct_perf}
\end{subfigure}
\hspace*{0.3cm}
\begin{subfigure}[t]{0.33\textwidth}
\includegraphics[width=\textwidth]{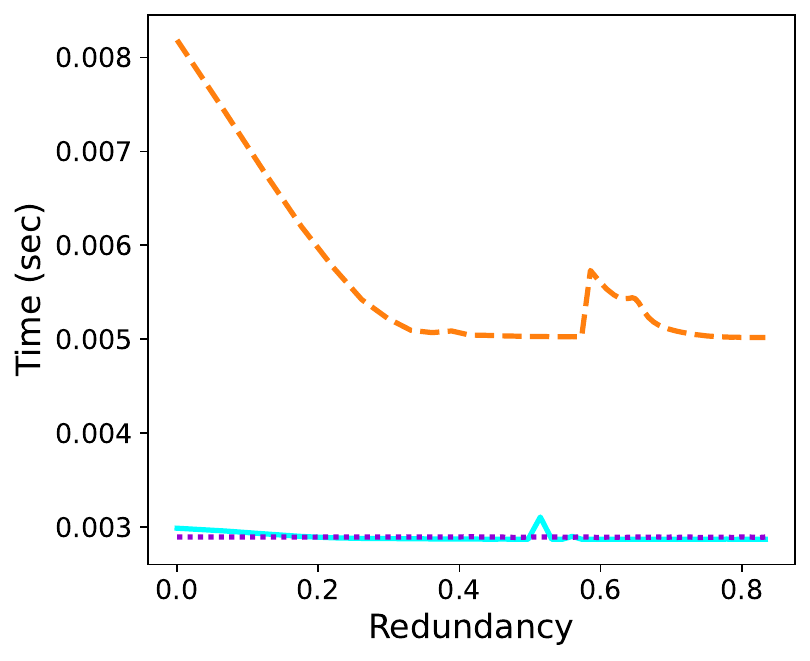}
\caption{Iterator}\label{fig:iter}
\end{subfigure}
\begin{subfigure}[t]{0.12\textwidth}
\vspace*{-3cm}
\includegraphics[width=\textwidth]{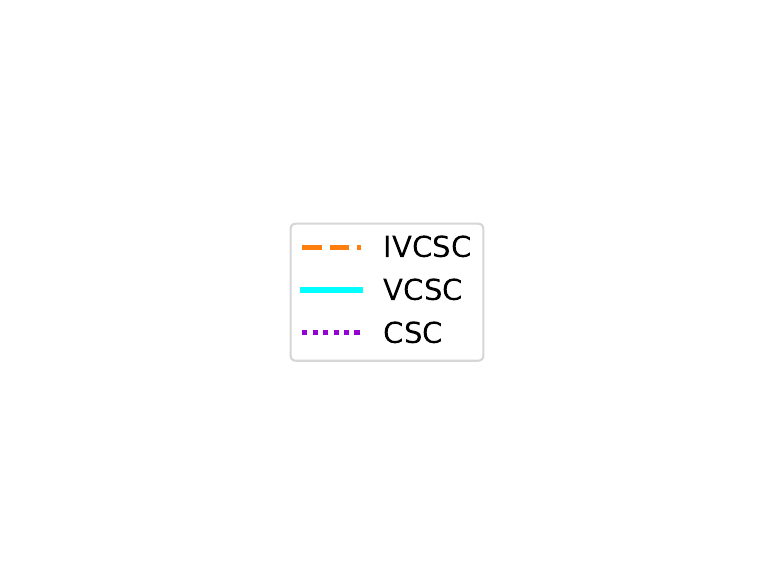}
\end{subfigure}

\vspace*{6pt}

\begin{subfigure}[b]{0.33\textwidth}
    \includegraphics[width=\textwidth]{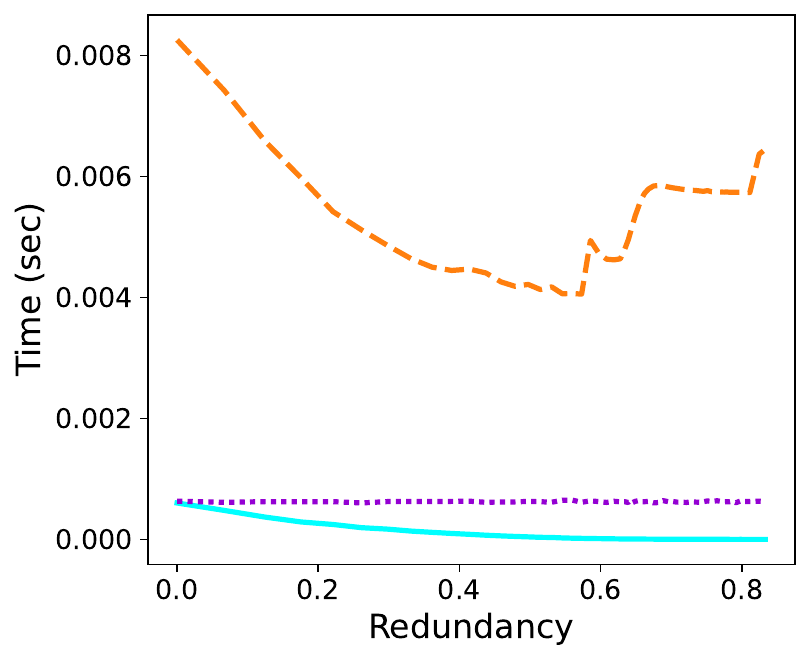}
    \caption{Scalar multiplication}\label{fig:scalar} 
\end{subfigure}
\centering
\begin{subfigure}[b]{0.33\textwidth}    \includegraphics[width=\textwidth]{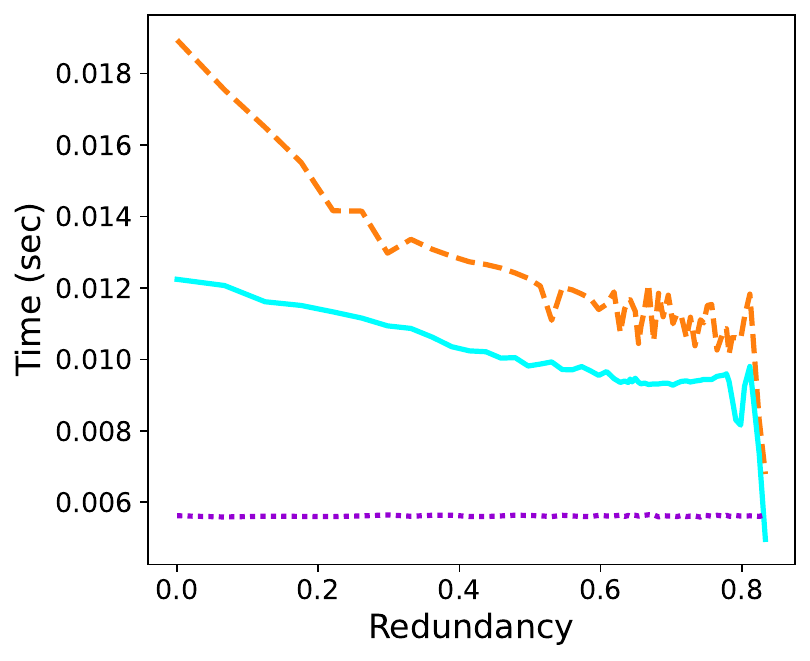}
    \caption{SpMV}
    \label{fig:spmv}
\centering
\end{subfigure}
{\begin{subfigure}[b]{0.31\textwidth}
    \includegraphics[width=\textwidth]{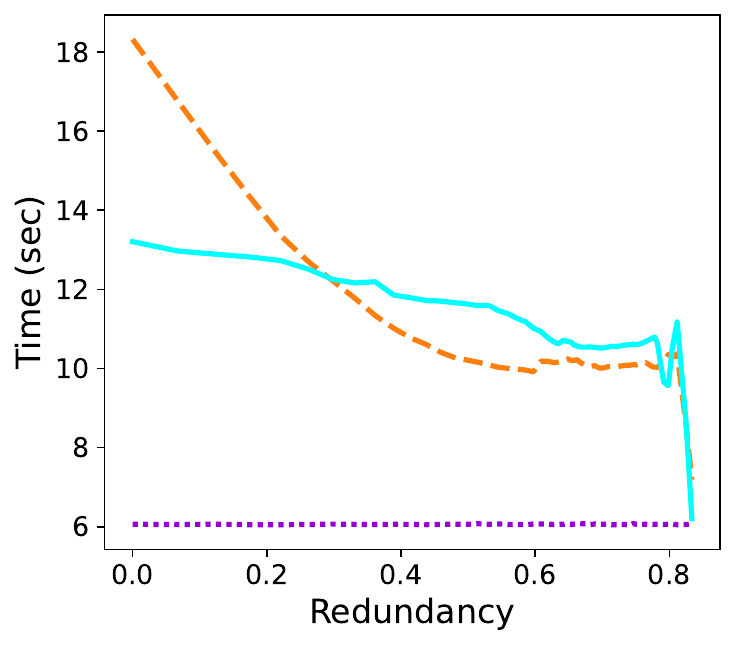}
    \caption{SpMM}
    \label{fig:spmm}
\end{subfigure}}
\caption{Benchmarking results for common operations on a randomly generated 90\% sparse $1 \text{ million} \times 25$ matrix, with values stored as floats (4 bytes) and indices stored as 4 byte integers.}
\label{fig:perf_plot}
\end{figure*}

Computational performance benchmarking results are shown in \autoref{fig:perf_plot}.  Constructor time is measured as the time to construct from a COO matrix.  As shown in \autoref{fig:construct_perf}, at low MMR, construction time is significantly slower for both VCSC and IVCSC, but it approaches the construction time of CSC at high MMR.  However, in absolute terms, the construction time is a one time cost that is typically negligible compared to repeated computations. 

Iterator traversal time is measured as the time necessary to fully iterate through the sparse matrix.
As shown in \autoref{fig:iter}, VCSC iterator traversal time is comparable to CSC iterator time, whereas IVCSC is 1.5x-3x slower.

Because scalar multiplication needs to loop over only the unique values in VCSC, for redundant matrices, element-wise operations are performed more quickly than in CSC, as is shown in \autoref{fig:scalar}.
In IVCSC, the values are not stored separately, thus necessitating iterating over all values and indices.  The upticks at higher redundancies correspond to cache effects due to changes in byte-packing.  This is not present in the iterator benchmarking, as the iterator benchmarking writes to only \texttt{volatile} memory, which is not cached.
For sparse matrix-vector (SpMV) and sparse matrix-matrix (SpMM), shown in \autoref{fig:spmv} and \autoref{fig:spmm}, respectively, VCSC (IVCSC) is approximately 2-fold (3-fold) slower for low MMR, dropping to roughly equivalent (marginally slower) for a fully redundant matrix.
IVCSC performing similarly to, or better than, VCSC on SpMM for high redundancies is reasonably consistent in our testing on large, simulated matrices and most likely due to cache effects.

\section{Conclusion and Future Work}
In this paper we presented two novel compression formats, VCSC and IVCSC, for matrices with highly redundant values.
Benchmarking showed that both VCSC and IVCSC offer considerable compression over CSC on data with high MMR, with a reasonably small performance hit.
One disadvantage of IVCSC is the slower iterator traversal, which results in increased compute time.
However, the up to 1.9-fold and 4.4-fold decrease in size for VCSC and IVCSC when compared to CSC, respectively, may allow in-core processing of very large matrices that could exhaust RAM on typical workstations, particularly as the size of datasets continues to grow.

This work lays the foundation for future research on hybrid sparse matrix formats, such as a hybrid CSC-VCSC structure that uses VCSC to store redundant values and CSC to store non-redundant values for any given column.
Additionally, distributed memory and SIMD parallelization of both VCSC and IVCSC could be beneficial to large scale machine learning applications.

\section{Acknowledgements}

This work was funded by a grant from the Chan Zuckerberg Initiative Single Cell Biology Data Insights DI-000-0287 (to Z.D., T.T., S.R., S.W.) and a Grand Valley State University Kindschi Fellowship (to S.W.).  This research made use of Clipper, the high-performance computing cluster at Grand Valley State University.


\bibliographystyle{includes/IEEEbib}
\bibliography{vcsc}

\end{document}